%% file: main.tex
\documentclass[sigconf]{acmart}
\AtBeginDocument{%
  \providecommand\BibTeX{{%
    \normalfont B\kern-0.5em{\scshape i\kern-0.25em b}\kern-0.8em\TeX}}}

\setcopyright{acmcopyright}
\copyrightyear{2018}
\acmYear{2018}
\acmDOI{10.1145/1122445.1122456}

\begin{document}



\title{Eight Reasons Why Cybersecurity on Novel Generations of Brain-Computer Interfaces Must Be Prioritized}

\author{Sergio López Bernal}
\email{slopez@um.es}
\orcid{0000-0003-1869-1965}
\affiliation{%
  \institution{Department of Information and Communications Engineering, University of Murcia}
  \city{Murcia}
  \country{Spain}
  \postcode{30100}
}

\author{Alberto Huertas Celdrán}
\email{huertas@ifi.uzh.ch}
\orcid{0000-0001-7125-1710}
\affiliation{%
  \institution{Communication Systems Group (CSG), Department of Informatics (IfI), University of Zurich UZH}
  \city{Zürich}
  \country{Switzerland}
  \postcode{8050}
}

\author{Gregorio Martínez Pérez}
\email{gregorio@um.es}
\orcid{0000-0001-5532-6604}
\affiliation{%
  \institution{Department of Information and Communications Engineering, University of Murcia}
  \city{Murcia}
  \country{Spain}
  \postcode{30100}
}

\renewcommand{\shortauthors}{López Bernal, et al.}

\ccsdesc[500]{Security and privacy~Domain-specific security and privacy architectures}

\keywords{Cybersecurity, Brain-Computer Interfaces, Neuronal Cyberattacks, Taxonomy}

\maketitle


\section{Introduction}
\label{sec:introduction}
\input{tex/1introduction.tex}

\section{The brain at risk due to novel generations of BCI}
\label{sec:background}
\input{tex/2background.tex}

\section{Eight Neural Cyberattacks affecting brain behavior}
\label{sec:taxonomy}
\input{tex/3taxonomy.tex}

\section{What is the impact of neural cyberattacks?}
\label{sec:experimentation}
\input{tex/4results.tex}

\section{Conclusion}
\label{sec:conclusion}
\input{tex/5conclusion.tex}

\begin{acks}
This work has been partially supported by (a) Bit \& Brain Technologies S.L. under the project CyberBrain, associated with the University of Murcia (Spain), by (b) the Swiss Federal Office for Defense Procurement (armasuisse) with the CyberSpec (CYD-C-2020003) project, and by (c) the University of Zürich UZH. We thank Blausen Medical\footnote{https://doi.org/10.15347/wjm/2014.010} and Harryarts\footnote{https://www.freepik.com/free-vector/lineal-brain-design\_841425.htm} from their publicly available images.
\end{acks}



\bibliographystyle{ACM-Reference-Format} 
\bibliography{references}

\end{document}

%% file: tex/1introduction.tex
Brain-Computer Interfaces (BCIs) are bidirectional systems that interact with the brain, allowing the acquisition of neural data and neuronal stimulation. BCIs can be classified according to their invasiveness level, being invasive interfaces extensively used in medical therapy. In this context and as an example, invasive BCIs focused on neural recording have been used to control prosthetic limbs in impaired patients, while BCIs for neuromodulation have been helpful for treating neurodegenerative conditions, such as Parkinson's disease \cite{Hartmann:parkinson:2019}. The second main family of BCIs, in terms of invasiveness, is the non-invasive one. BCIs based on non-invasive principles and, mainly, those focused on neural data acquisition such as electroencephalography (EEG), have gained popularity in recent years, extending their usage from traditional medical scenarios to new domains such as entertainment or video games. However, despite the benefits of non-invasive BCIs, some works in the literature have identified particular cybersecurity issues from a neural data acquisition perspective. In particular, Martinovic et al. \cite{Martinovic:visual:2012} demonstrated that an attacker could obtain sensitive personal data from BCI users, taking advantage of their cerebral response (P300 potentials) generated when known visual stimuli are presented to them. Bonaci et al. \cite{Bonaci:appStores:2015} also described a scenario where attackers could maliciously add or modify software modules defining the BCI to perform dangerous actions over the users. Finally, Takabi et al. \cite{Takabi:privacyThreatsCounter:2016} highlighted that most APIs used to develop BCI applications offered complete access over the information acquired by the BCI, presenting confidentiality problems.

Cybersecurity of invasive BCIs is also a challenge that has been identified in the literature and whose application is in initial stages. This situation is complicated by the recent introduction of novel BCI designs based on nanotechnology aiming to surpass the limitations of traditional BCIs. One example of these emergent systems is Neuralink \cite{Neuralink}, which uses nanotechnology to record and stimulate particular brain regions with single-neuron resolution. Despite the advantages of the new generation of invasive BCIs, the literature has already identified that some of these BCIs present vulnerabilities that attackers could exploit to affect neural activity \cite{Lopez_Bernal:cyberattacks_implants:2020}. In particular, the literature has proposed two cyberattacks focused on neural stimulation named Neural Flooding and Neural Scanning \cite{Lopez_Bernal:cyberattacks_implants:2020}, as well as a cyberattack focused on neural inhibition \cite{arxiv_COSE}. These threats have been defined within the term \textit{neural cyberattacks}, consisting in well-known attacks from computer able, able to disrupt the spontaneous activity of neural networks of the brain, stimulating or inhibiting neurons.

In such a disruptive and novel context, one of the main challenges is formally defining the behavior of different neural cyberattacks affecting the brain. In this direction, studies addressing how neural cyberattacks could recreate the effects induced by certain neurodegenerative diseases are absent in current literature. Furthermore, the analysis of these cyberattacks regarding their impact on spontaneous neural activity is unexplored. Finally, a comparison of the impact caused by distinct neural cyberattacks is required to understand the changes caused over the brain.

With the goal of improving the previous open challenges, this article presents eight neural cyberattacks affecting spontaneous neural activity, inspired by well-known cyberattacks from the computer science domain: Neural Flooding, Neural Jamming, Neural Scanning, Neural Selective Forwarding, Neural Spoofing, Neural Sybil, Neural Sinkhole and Neural Nonce. After presenting their formal definitions, the cyberattacks have been implemented over a Convolutional Neural Network (CNN) simulating a portion of a mouse's visual cortex. This implementation is based on existing works indicating the similarities that CNNs have with neuronal structures from the visual cortex \cite{Gal2017}. Finally, a comparison of the impact between each neural cyberattack is presented for the initial and final part of a neural simulation, studying their impact for both the short and long term. In conclusion, Neural Nonce and Neural Jamming are the most suitable cyberattacks for short-term effects, while Neural Scanning and Neural Nonce are the most adequate for long-term effects.

%% file: tex/2background.tex
Although this work focuses on neuronal cyberattacks from a Computer Science point of view, it is essential to introduce, in a basic and synthesized way, how the brain works to understand their behavior and the current state of neuromodulation technologies able to stimulate and/or inhibit neurons.

The brain is the most complex organ in the human body, managing all major activities of the organism. Its structure is divided into two hemispheres, left and right, controlling the opposite side of the body. Moreover, the cortex of each hemisphere presents four lobes on its surface with differentiated responsibilities. Frontal lobes intervene in reasoning, planning, translating thoughts into words, and defining personality. In contrast, parietal lobes manage sensory perceptions such as taste or touch, additionally to temperature and pain. These lobes also intervene in memory and the understanding of languages. Occipital lobes are in charge of decoding visual information, such as colors or forms, and identify objects, while temporal lobes focus on processing auditory stimuli, also intervening in verbal memory \cite{Kandel:NeuralScience:2013}. 

Within the hemispheres, around 86 billion neurons interact with each other to perform these complex tasks. This interaction is performed by two specific structures of the neuron, the dendrites and the axon. While dendrites receive information from other neurons, axons transmit instructions to neurons. The connection established between these structures is known as a synapse, and it is the base of neuronal communication. In neuronal communication, the dendrites of a given neuron receive stimuli from many neurons (presynaptic neurons) via neurotransmitters, which are molecules that force actions in the receiver neuron (postsynaptic neuron). Presynaptic neurons can be excitatory, producing particular neurotransmitters aiming to initiate an impulse on the postsynaptic neuron or inhibitory, liberating neurotransmitters to prevent its activity. If the sum of these positive and negative impulses exceeds the excitation threshold of the postsynaptic neuron, this neuron will generate a nerve impulse known as action potential (or spike), electrically transmitted along the axon to reach the axon terminals. When the electric stimulus reaches these terminals, they liberate particular neurotransmitters to the synaptic gap, the space separating the axon from the dendrites of other neurons, aiming to influence their activity in an excitatory or inhibitory way. These electric and chemical processes are repeated neuron after neuron, only if they exceed their excitation threshold.

Neurotechnology plays an essential role in supporting these neuronal communications, used for decades in clinical scenarios to induce or suppress neural activity. There is a wide variety of technologies, both invasive and non-invasive, with different modulation principles such as ultrasounds, electrical currents, magnetic fields, or light pulses (optogenetics) \cite{Edwards2017}. Despite the differences of these approaches, most of them share common parameters used to adjust the modulation process, such as the amplitude or voltage applied or the duration and periodicity of the pulses. Focusing on invasive BCIs, Deep Brain Stimulation (DBS) represents an excellent example of these technologies used to treat conditions like Parkinson's disease or obsessive-compulsive disorder using neural stimulation \cite{Hartmann:parkinson:2019}. Moreover, most invasive BCIs also offer recording capabilities, enabling the monitoring of the brain to determine the best instant to stimulate or inhibit a particular set of neurons.

\begin{figure}[h]
\begin{center}
\includegraphics[width=\columnwidth]{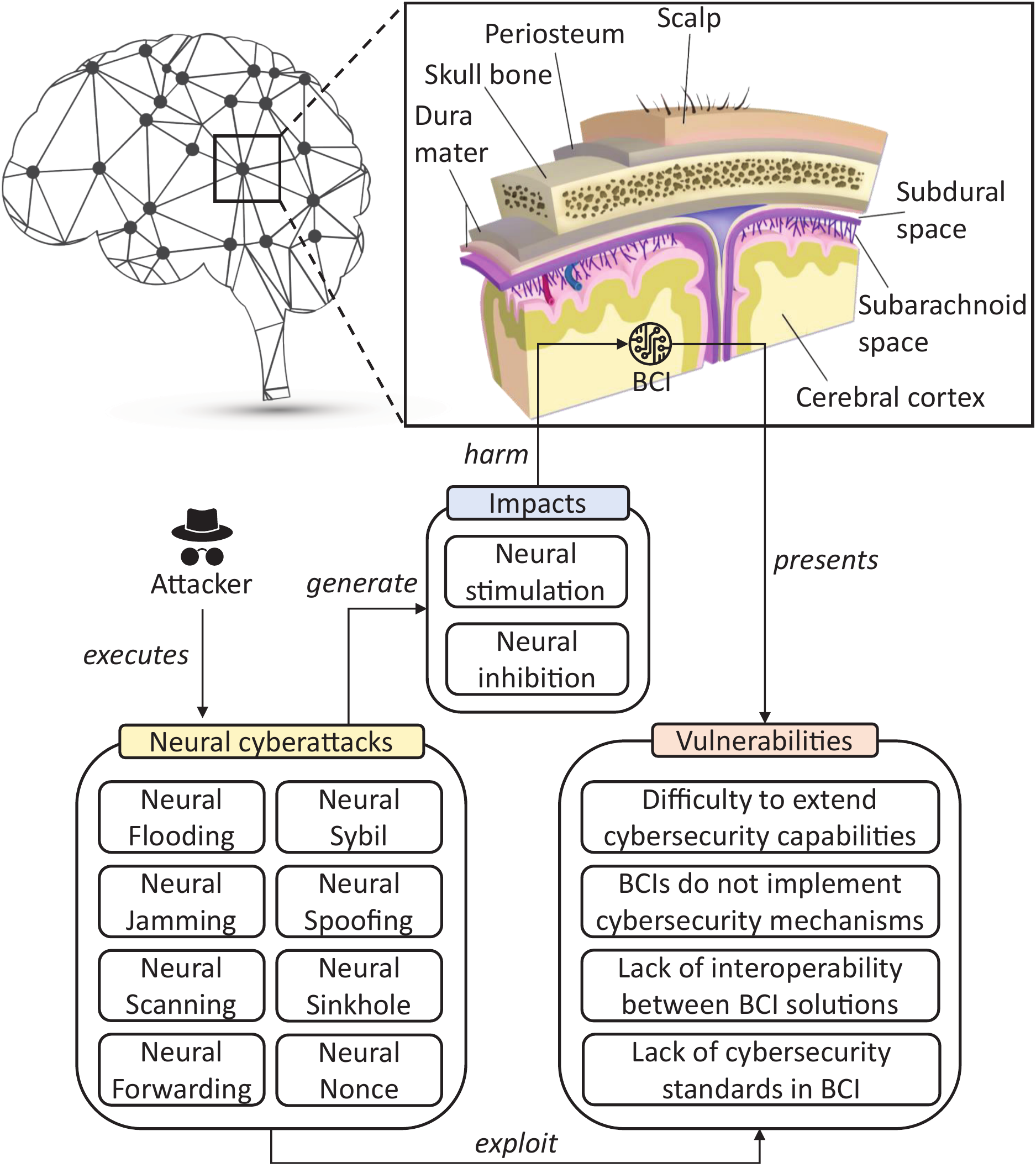}
\end{center}
\caption{Attacker executing the proposed neuronal cyberattacks that exploit vulnerabilities of invasive neuromodulation BCIs and generate particular impacts on the BCI.}
\label{fig:background}
\end{figure}

In such a scenario, novel solutions such as Neuralink \cite{Neuralink} or WiOptND \cite{Wirdatmadja2017} deserve special interest since they have introduced the use of nanotechnology to miniaturize the electrodes implanted in the brain, achieving single-neuron resolution. Particularly, these technologies address neuromodulation from two different perspectives. Neuralink uses electrical currents to stimulate the brain, while WiOptND stimulates or inhibits neuronal activity using optogenetics. Nevertheless, these current initiatives present vulnerabilities in their architectures that attackers could exploit to stimulate or inhibit neurons maliciously \cite{Lopez_Bernal:cyberattacks_implants:2020}. In this direction, \figurename~\ref{fig:background} introduces the anatomical structure of the head from the scalp to the cerebral cortex, presenting an invasive neuromodulation BCI placed in the cortex that an attacker externally targets. As can be seen, the attacker can execute one of the eight cyberattacks proposed in this work (more details are provided in Section \ref{sec:taxonomy}). These cyberattacks exploit vulnerabilities existing in current BCIs (see \cite{Lopez_Bernal:cyberBCI:2021}), generating an impact over the BCI, thus stimulating or inhibiting neuronal activity.

%% file: tex/3taxonomy.tex
Once the vulnerabilities of novel BCIs have been highlighted, it is time to introduce neural cyberattacks exploiting them and affecting neural behavior. In this direction, this work presents eight cyberattacks inspired by well-known threats from digital communications, justified by the potential exploitation of previously highlighted vulnerabilities. Particularly, five of these cyberattacks are new (Neural Selective Forwarding, Neural Spoofing, Neural Sybil, Neural Sinkhole and Neural Nonce, while the remaining three were presented in previous work (Neural Flooding and Neural Scanning in \cite{Lopez_Bernal:cyberattacks_implants:2020}, and Neural Jamming in \cite{arxiv_COSE}). All these cyberattacks are either based on the stimulation of neurons, their inhibition, or a combination of both. Particularly, and for the sake of simplicity, these cyberattacks assume the usage of technologies able to stimulate or inhibit neuronal behavior.



\subsection{Neuronal Flooding}
\label{subsec:taxonomy_FLO}
In cybersecurity, flooding cyberattacks focus on collapsing a network by transmitting a high number of data packets, generally directed to particular targets within the network. As a consequence, these endpoints increase their workload, not being able to manage legit communications adequately. Moving to a neurological perspective, Neuronal Flooding (FLO) cyberattacks aim to overstimulate multiple neurons in a particular time instant. This cyberattack does not need previous knowledge about the status of the target neurons, presenting a low complexity compared to other neural cyberattacks. 

\begin{figure}[h]
\begin{center}
\includegraphics[width=0.9\columnwidth]{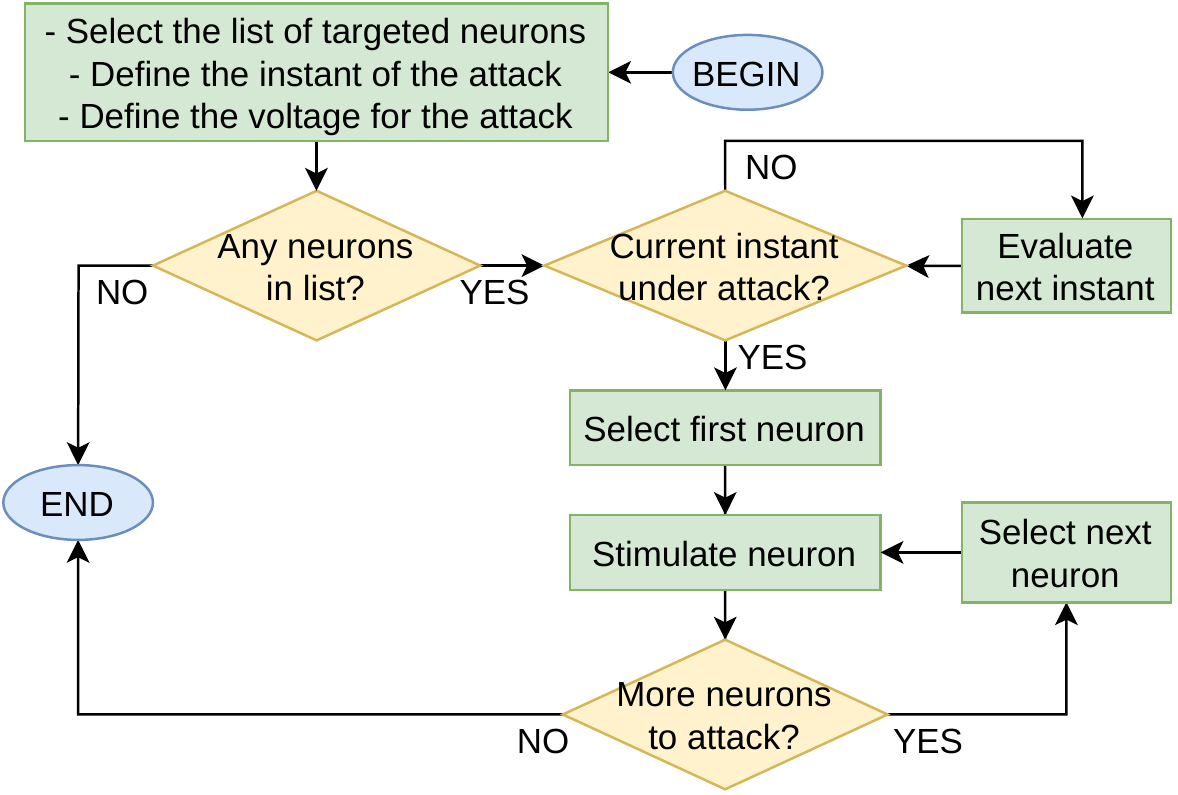}
\end{center}
\caption{Implemented behavior of Neuronal Flooding.}
\label{fig:flowchart_FLO}
\end{figure}

The general behavior of the FLO cyberattack implemented can be consulted in \figurename~\ref{fig:flowchart_FLO}, where green boxes indicate actions performed by the cyberattack, and yellow diamonds are conditional blocks. First, the attacker determines the attacking instant and the list of targeted neurons. During the desired instant, the cyberattack selects each of the neurons and stimulates it. Although the flow chart presented could be interpreted as sequentially affecting these neurons, the attack is performed in a particular instant of time, resulting in attacking the neurons at the same time. 

\subsection{Neuronal Jamming}
\label{subsec:taxonomy_JAM}
Jamming cyberattacks focus on disrupting legitimate communications by introducing a malicious interference to the medium and preventing the devices from communicating, thus deriving in a denial of service (DoS). This principle can be translated to the neurological world, where Neuronal Jamming (JAM) consists in the inhibition of the activity of a set of neurons, impeding them from generating or transmitting impulses to adjacent neurons. In contrast to FLO, this cyberattack is performed during a determined temporal window, in which the affected neurons do not generate activity. This cyberattack also presents a low execution complexity, only requiring selecting the target neurons and the attack duration.

\begin{figure}[h]
\begin{center}
\includegraphics[width=\columnwidth]{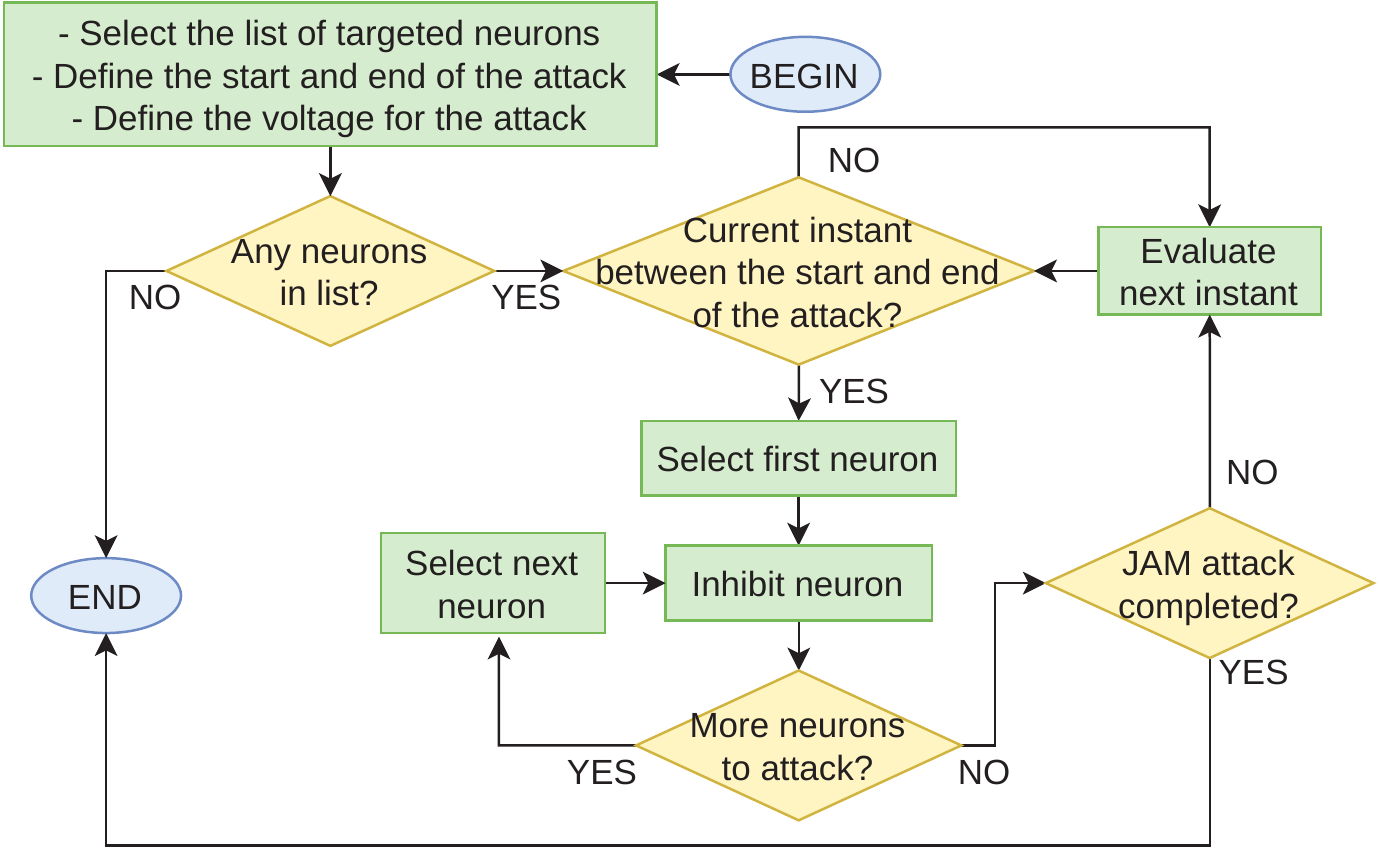}
\end{center}
\caption{Implemented behavior of Neuronal Jamming.}
\label{fig:flowchart_JAM}
\end{figure}

The flow chart depicted in \figurename~\ref{fig:flowchart_JAM} represents a temporal window in which the JAM cyberattack is performed. For each instant between the beginning and end of the attack, the list of targeted neurons is simultaneously inhibited. This inhibition consists in setting the neurons to their lowest voltage within their natural range of values. 

\subsection{Neuronal Scanning}
\label{subsec:taxonomy_SCA}
Port scanning is a common cybersecurity technique used to verify if the communication ports of a machine are being used and identify vulnerable services available in those ports. For that, all ports of the machine are sequentially tested. Similarly, Neuronal Scanning (SCA) cyberattacks aim to sequentially stimulate all neurons of a neuronal population, affecting only one neuron per time instant. As in the previous cyberattacks, SCA does not require previous knowledge about the status of the targeted neurons. Nevertheless, it presents a moderate execution complexity since the attacker needs to coordinate the order of the neurons attacked, avoid repetitions between them, and determine the time interval between attacking each neuron.

\begin{figure}[h]
\begin{center}
\includegraphics[width=0.9\columnwidth]{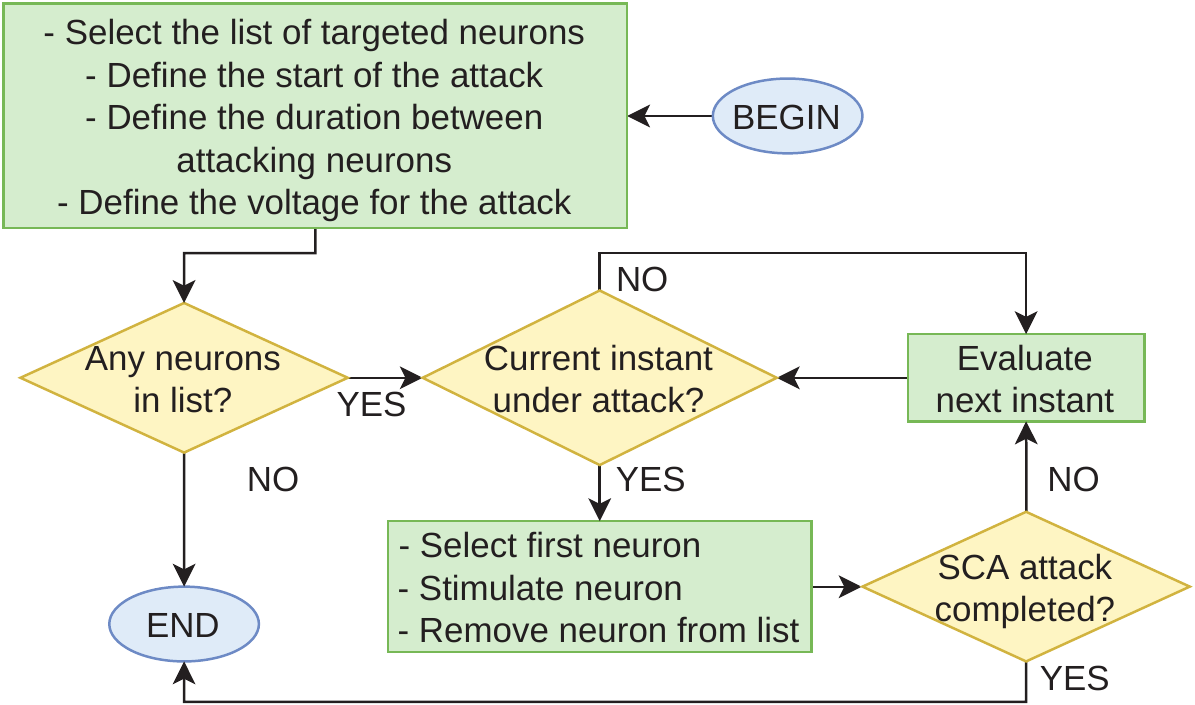}
\end{center}
\caption{Implemented behavior of Neuronal Scanning.}
\label{fig:flowchart_SCA}
\end{figure}

The SCA cyberattack implemented (see \figurename~\ref{fig:flowchart_SCA}) targets one neuron per instant under attack, removing from the list those neurons already attacked to avoid repetitions and ensure a sequential selection. These instants are determined based on the start of the attack and the time that the attacker waits between affecting neurons. 

\subsection{Neuronal Selective Forwarding}
\label{subsec:taxonomy_FOR}
Selective forwarding is one of the most harmful cyberattacks against communication networks. In this kind of threat, malicious hosts selectively drop some packets instead of forwarding them. The selection of dropping nodes may be random or predefined depending on the attack design. In the brain context, Neuronal Selective Forwarding (FOR) consists in changing the propagation behavior of a set of neurons during a temporal window, inhibiting particular neurons at each instant of the window. This attack is more elaborate than the previous ones because it requires knowledge of the neurons involved in a given neuronal propagation path and their status in each instant. It is achieved by real-time neuronal monitoring or previously knowing the neuronal propagation behavior due to the repetition of actions such as eye blinks or limb movements. 

\begin{figure}[h]
\begin{center}
\includegraphics[width=\columnwidth]{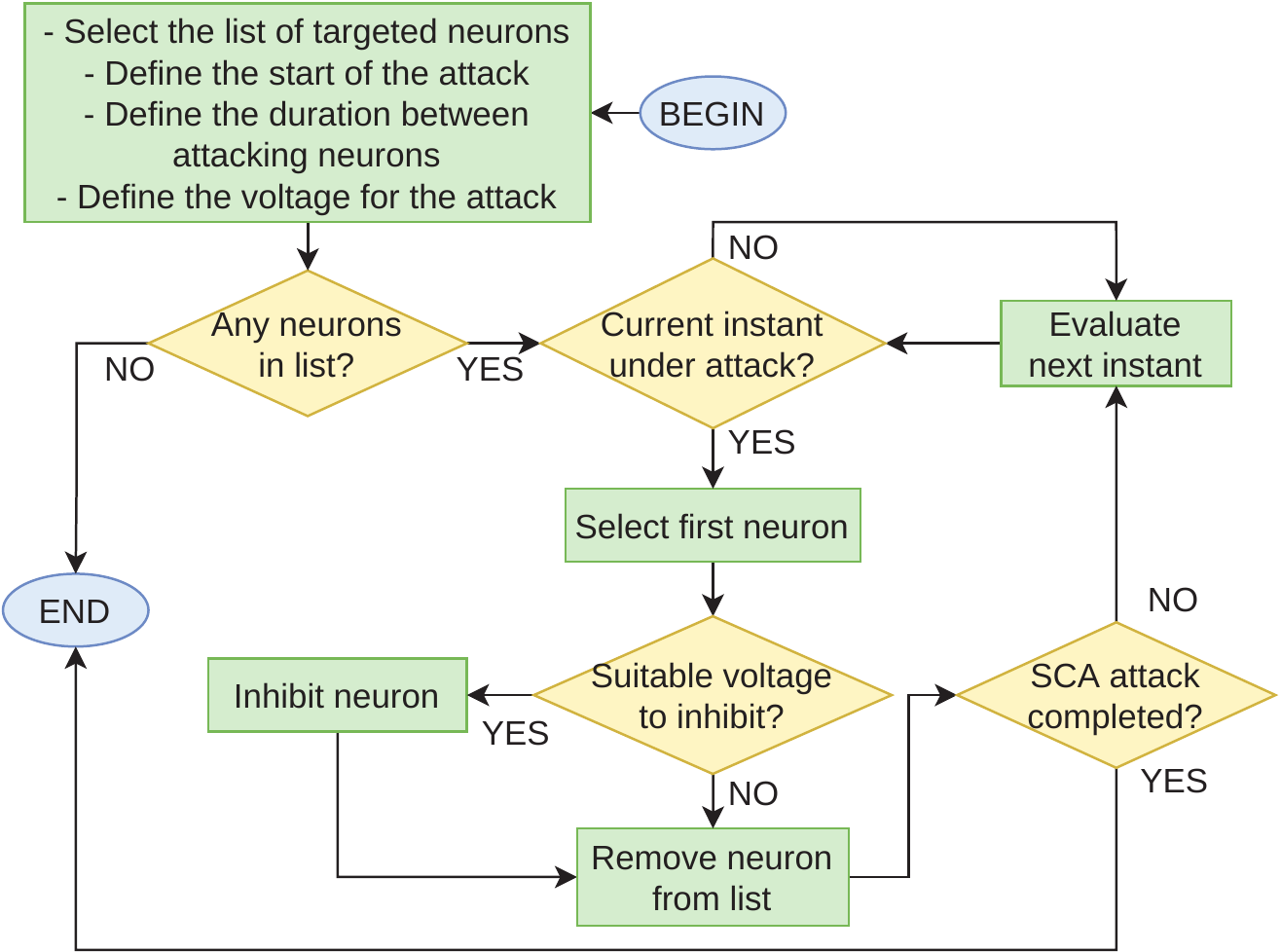}
\end{center}
\caption{Implemented behavior of Neuronal Selective Forwarding.}
\label{fig:flowchart_FOR}
\end{figure}

This cyberattack allows a wide variety of different configurations for targeting neurons. It has been followed the same sequential criteria already presented for SCA in this work, inhibiting them instead of performing neural stimulation. Attending to \figurename~\ref{fig:flowchart_FOR}, FOR introduces an additional conditional block that verifies if the current voltage of the neuron is suitable for inhibition. Based on the voltage defined for the attack, the implementation verifies if the subtraction between the current voltage and the attacking voltage is lower than the lowest possible value. If so, the attack sets the voltage to the lowest threshold to avoid unrealistic results.

\subsection{Neuronal Spoofing}
\label{subsec:taxonomy_SPO}
In computer networks, a spoofing cyberattack occurs when a malicious party impersonates a computer or subject to steal sensitive data or launch attacks against other network hosts. In the brain scenario, Neuronal Spoofing (SPO) cyberattacks consist in replicating the behavior of a set of neurons during a given period. After recording the neuronal activity, the attacker uses this pattern to stimulate or inhibit the same or different neurons at a different time. This attack is one of the most sophisticated since it requires recording, stimulation and inhibition capabilities and deep knowledge of brain functioning. Like most of them, the impact of this cyberattack is high because a malicious attacker could control some vital functions of the subject's body.

\begin{figure}[h]
\begin{center}
\includegraphics[width=\columnwidth]{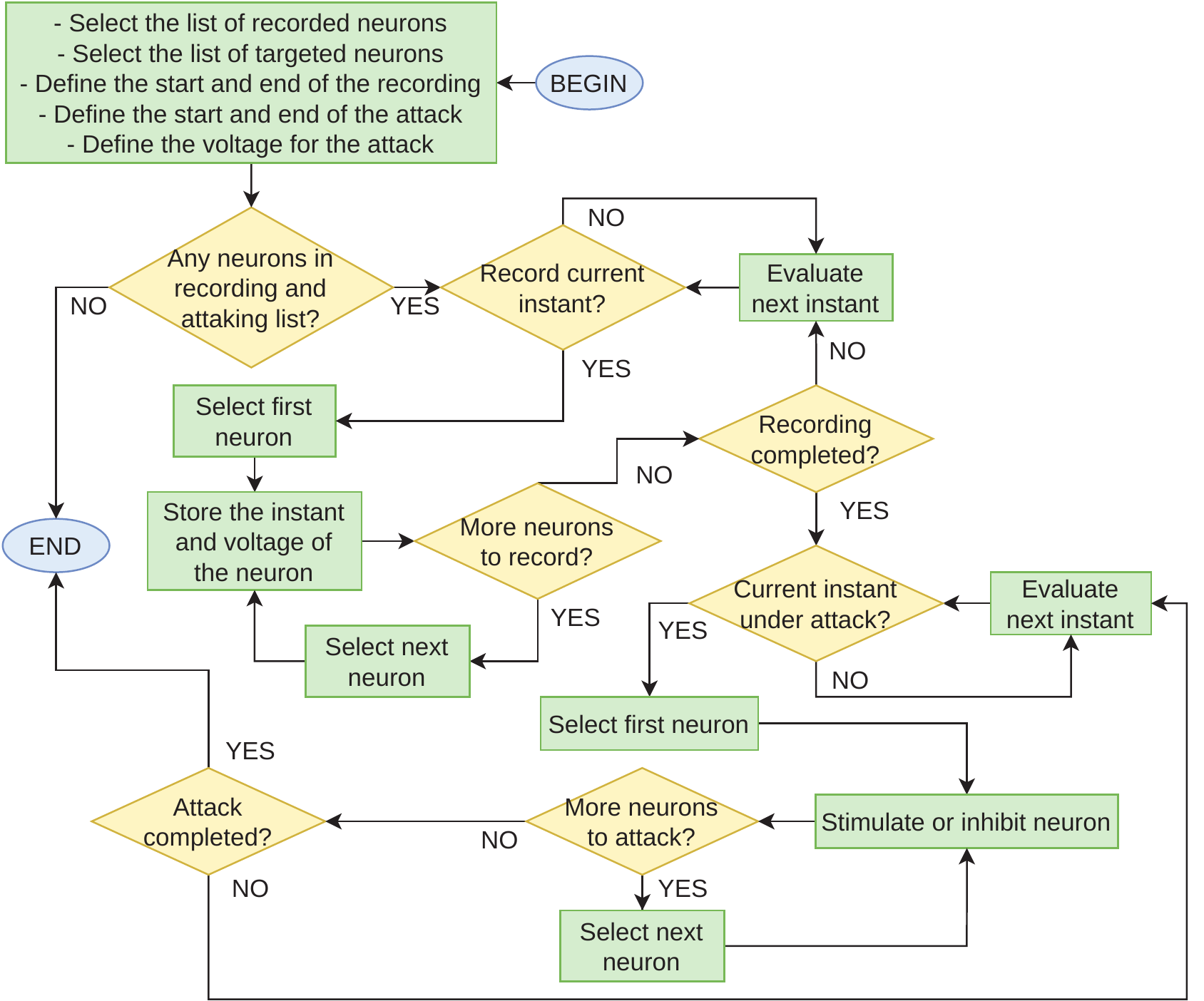}
\end{center}
\caption{Implemented behavior of Neuronal Spoofing.}
\label{fig:flowchart_SPO}
\end{figure}

The diagram presented in \figurename~\ref{fig:flowchart_SPO} highlights two main processes. First, the attack performs a neuronal recording procedure for the selected neurons during a particular temporal period. For each instant within the period, the attacker stores the voltage of each recorded neuron. After that, the second process properly stimulates or inhibits a different neuronal population targeted by the attack, forcing them to have the same behavior that those previously recorded. 

\subsection{Neuronal Sybil}
\label{subsec:taxonomy_SYB}
Sybil cyberattacks happen when a computer is hijacked to claim multiple identities, presenting broad security and safety implications. Having different identities, the behavior of the infected host differs according to the identity acting in each moment. Bringing these cyberattacks to the brain scenario implies that an attacker could alter the operation of one or more neurons doing precisely the opposite as their natural behavior. It means that when a given neuron is firing, the attacker inhibits the activity, and when it is not firing, the attacker fires it. Neuronal Sybil (SYB) cyberattacks are the most complex of the presented because they require real-time recording (or previous knowledge of the firing pattern) and the functionality of either stimulating or inhibiting a particular neuron in a given instant depending on its natural behavior. The impact of these neural cyberattacks is high, depending on the number of affected neurons. 

\begin{figure}[h]
\begin{center}
\includegraphics[width=0.9\columnwidth]{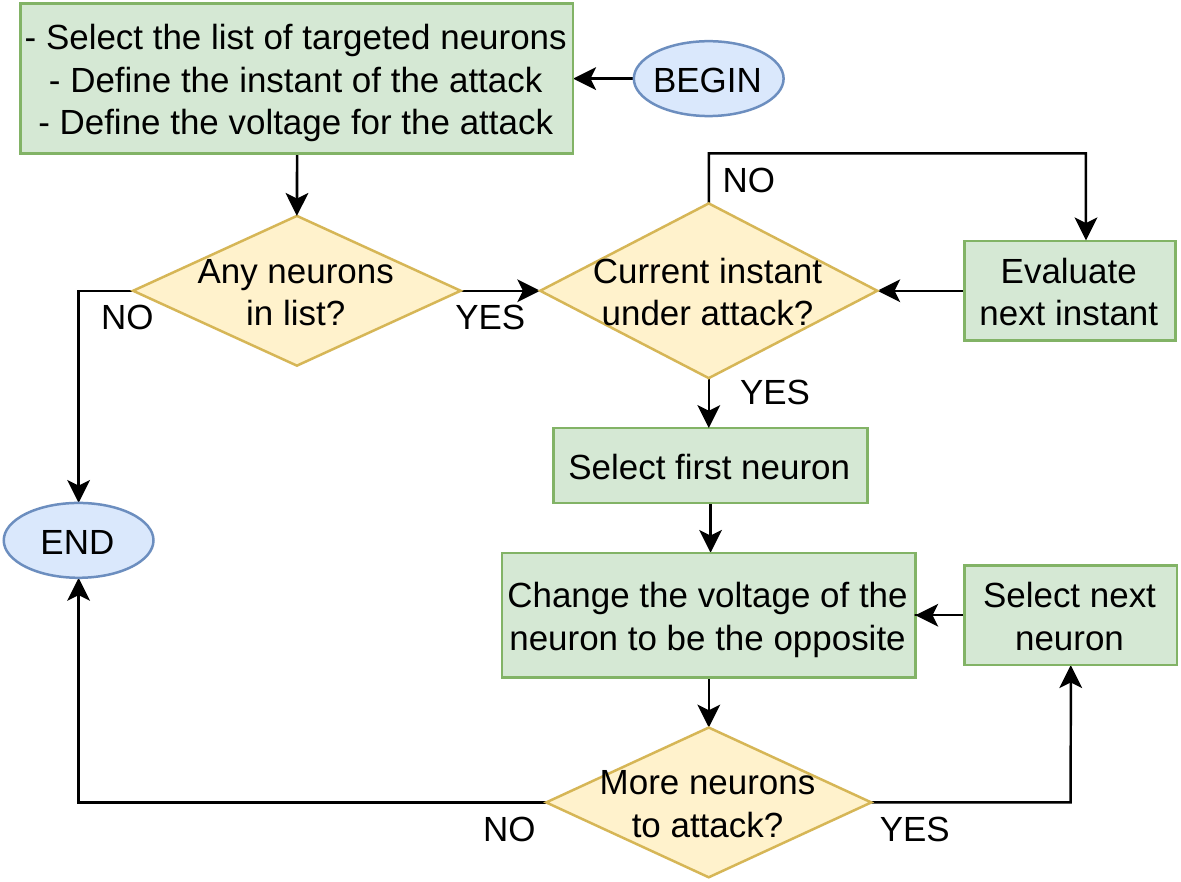}
\end{center}
\caption{Implemented behavior of Neuronal Sybil.}
\label{fig:flowchart_SYB}
\end{figure}

The implementation of SYB cyberattacks is similar to the presented for FLO, although the action performed over the neurons is different (see \figurename~\ref{fig:flowchart_SYB}). In SYB, the voltage of each targeted neuron is set to the opposite value within its natural range. This is obtained by adding the higher and lower voltage thresholds of the neuron and subtracting the current voltage value. 

\subsection{Neuronal Sinkhole}
\label{subsec:taxonomy_SIN}
Sinkhole cyberattacks are applied to routing protocols, where a node of the network broadcasts that itself is the best path to reach particular destinations. Based on that, the surrounding nodes will transmit their traffic to the malicious node, which could access, modify or discard the received data. From a neurological perspective, Neuronal Sinkhole (SIN) cyberattacks focus on stimulating neurons from superficial layers connected to neurons placed in deeper layers, being the later the main target of the attack. In this regard, SIN cyberattacks present a high complexity since the attacker requires knowledge about the neuronal topology and synapses of a specific area of the brain. Moreover, this cyberattack is performed in a particular instant, stimulating the trigger set of neurons that initiates the attack. 

\begin{figure}[h]
\begin{center}
\includegraphics[width=0.9\columnwidth]{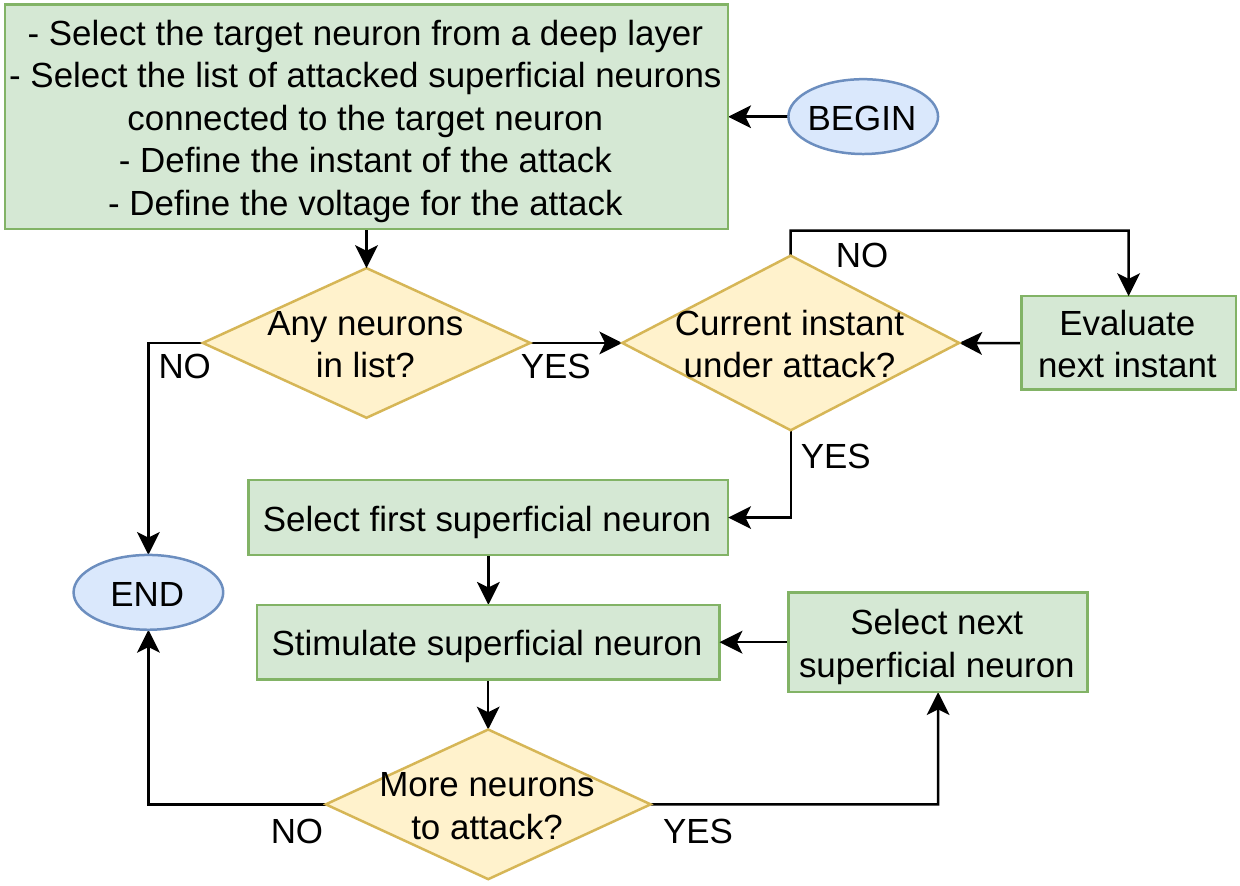}
\end{center}
\caption{Implemented behavior of Neuronal Sinkhole.}
\label{fig:flowchart_SIN}
\end{figure}

The actions included in the implementation of SIN cyberattacks are the same as the presented for FLO, as shown in \figurename~\ref{fig:flowchart_SIN}. The main difference between them lies in the selection of the targeted neurons. SIN cyberattacks directly affect the neurons from early layers connected via synapses with the target neuron located in deep layers. Once identified the neurons to attack, the process of stimulation is the same as FLO. 

\subsection{Neuronal Nonce}
\label{subsec:taxonomy_NON}
Nonce numbers are typically random values utilized in cryptography to secure communications. A nonce is typically used just once to prevent that old communications are reused and thus perform a replay attack. In the context of neural cyberattacks, Neuronal Nonce (NON) consists in attacking a random set of neurons in a particular instant of time. The action performed could vary based on the interests of the attacker, either producing neural stimulation, neural inhibition, or a combination of both. The next execution of the attack will target a completely different set. Based on this variability, the complexity of the cyberattack is low, just requiring physical access to the target neurons.

\begin{figure}[h]
\begin{center}
\includegraphics[width=\columnwidth]{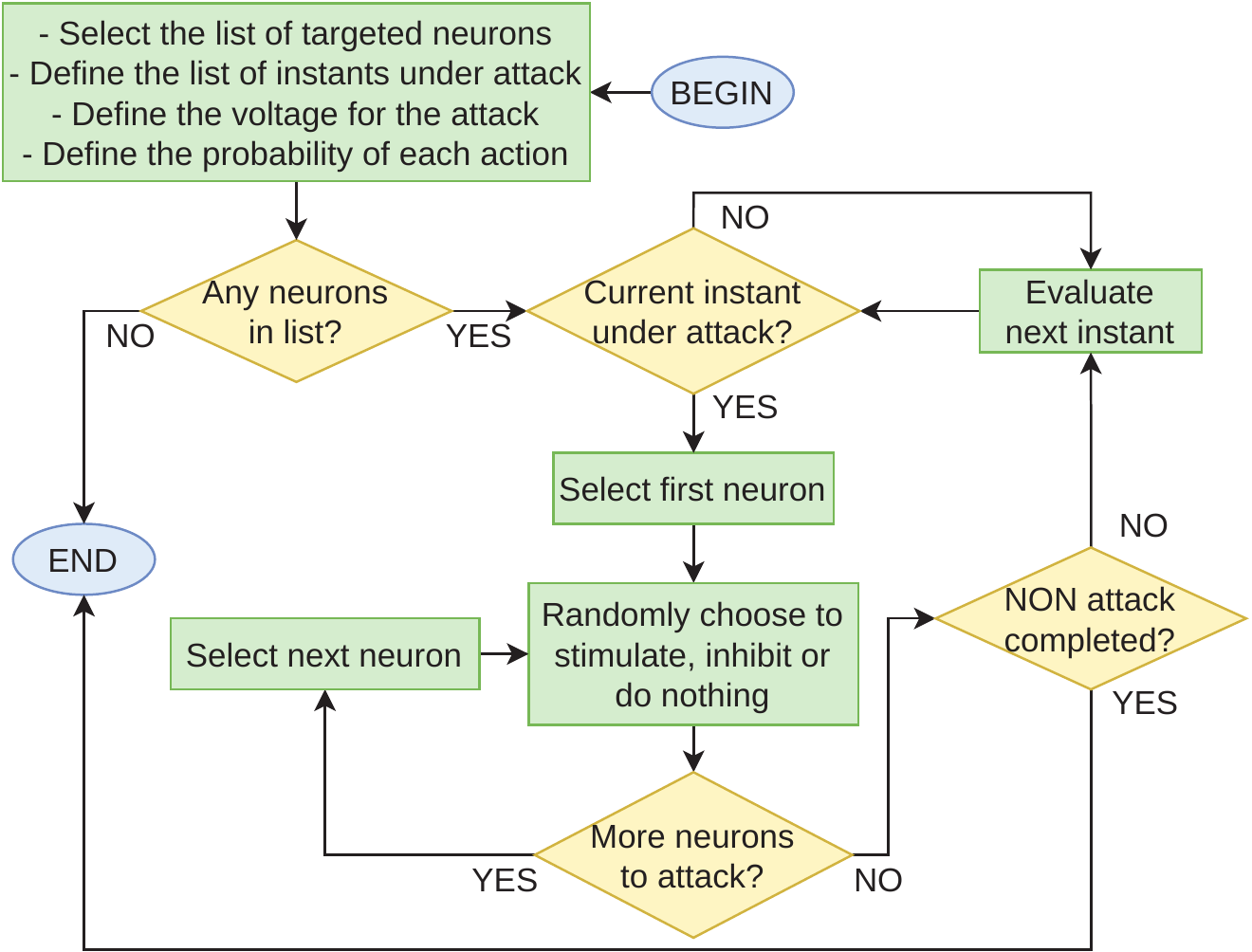}
\end{center}
\caption{Implemented behavior of Neuronal Nonce.}
\label{fig:flowchart_NON}
\end{figure}

This cyberattack has been implemented following the same principles already presented. The main difference (see \figurename~\ref{fig:flowchart_NON}) resides in the selection of the action to apply over each targeted neuron. For each instant under attack and each targeted neuron, the attack randomly determines to stimulate, inhibit or keep its spontaneous behavior. The attacker can also indicate the probability assigned to each action aiming to benefit particular actions. 

Once presented the behavior of each neural cyberattack, \tableautorefname~\ref{table:comparison_attacks} introduces a comparison between them. In particular, the theoretical impact of each attack depends on the aggressiveness of its action mechanism and the knowledge that the attacker has about the target neurons. Nevertheless, these cyberattacks present particular aspects that complicate their comparison, such as their inner behavior, the instants and duration of the cyberattacks, the number of affected neurons, or the voltages used to stimulate those neurons. 

\begin{table}[h]
\caption{Comparison of proposed neural cyberattacks}
\label{table:comparison_attacks}
\setlength\tabcolsep{2pt}
\centering
\resizebox{\columnwidth}{!}{
\begin{tabular}{|c|c|c|c|c|}
\hline

\textbf{Cyberattack} & \textbf{Impact} & \begin{tabular}{@{}c@{}} \textbf{Neurons involved} \\ \textbf{per instant} \end{tabular} & \textbf{Duration} & \textbf{Complexity} \\ \hline

Neuronal Flooding & Stimulation & 1 - many & One instant & Low \\ \hline
Neuronal Jamming & Inhibition & 1 - many & Time window & Low \\ \hline
Neuronal Scanning & Stimulation & 1 & Time window & Moderate \\ \hline
\begin{tabular}{@{}c@{}} Neuronal Selective \\ Forwarding \end{tabular} & \begin{tabular}{@{}c@{}} Recording \\ Inhibition \end{tabular} & 1 - many & Time window  & Moderate \\ \hline
Neuronal Spoofing & \begin{tabular}{@{}c@{}} Recording \\ Stimulation \\ Inhibition \end{tabular} & 1 - many & Time window & High  \\ \hline
Neuronal Sybil & \begin{tabular}{@{}c@{}} Recording \\ Stimulation \\ Inhibition \end{tabular} & 1 - many & One instant & High  \\ \hline
Neuronal Sinkhole & Stimulation & 1 - many & One instant & Low  \\ \hline
Neuronal Nonce & \begin{tabular}{@{}c@{}} Recording \\ Inhibition \end{tabular} & 1 - many & One instant & Low  \\ \hline
\end{tabular}}
\end{table}

%% file: tex/4results.tex
To answer this question it is important to mention that Neural topologies are critical to measure the impact caused by cyberattacks, and there is an absence of realistic neuronal topologies in the literature, being an open challenge of the area \cite{Gal2017}. In this context and with the goal of alleviating this limitation, the literature has evidenced that the hierarchy and functioning of neurons in charge of the vision present similarities with the functioning of CNNs \cite{Kuzovkin2018}. Particularly, the layers in both networks move from simple to abstract, where convolutional layers are related to early visual regions and dense layers present similarities with later visual areas. Based on that, this work employs a neuronal topology artificially generated from training a CNN, where the resulting weights are transformed to synaptic weights, used to represent the voltage increase induced during an action potential. 

Considering the similarities between CNNs and biological approaches, previous work trained a CNN to solve the problem of a mouse trying to exit a determined maze, modeling a portion of a mouse's visual cortex \cite{Lopez_Bernal:cyberattacks_implants:2020, arxiv_COSE}. This CNN was trained to obtain the optimal path on the maze to find the exit, resulting in 27 positions, whose topology comprises two convolutional layers of 200 and 72 nodes, respectively, and a final dense layer of four nodes. Although this topology is not equivalent to a biological one, it serves to compare the impact that each neural cyberattack has over a common baseline.

Once having the neural topology, it was ported to the Brian2 neuronal simulator \cite{Stimberg2019}, modeling the behavior of pyramidal neurons from three different layers of the visual cortex of the mouse (L2/3, L5, and L6). For that, Izhikevich's neuronal model \cite{Izhikevich2003} was used to represent excitatory neurons with regular spiking dynamics, defining neurons with a voltage range between -65 mV and 30 mV. Finally, a simulation of 27 seconds was defined, simulating a mouse staying one second in each position of the optimal path of the maze previously mentioned. Supplementary information concerning design and implementation aspects can be found in \cite{Lopez_Bernal:cyberattacks_implants:2020}.

\tablename~\ref{table:parameters} summarizes the parameters used during the experimentation for each neural cyberattack. It is relevant to note that FLO, JAM, SPO, and SYB target random neurons from the first layer, while SCA and FOR sequentially attack all 200 neurons. SIN affects only the neurons related to the target neurons, and NON randomly evaluates the decision over each neuron of the first layer. Finally, NON presents a probability of 20\% of stimulating a neuron, a 20\% of inhibiting it, and a remaining 60\% of keeping its spontaneous behavior until the next instant under attack.

\begin{table}[h]
\caption{Parameters used for each neural cyberattack, where up arrows ($\uparrow$) indicate a voltage increase, and down arrows ($\downarrow$) a voltage decrease.}
\label{table:parameters}
\setlength\tabcolsep{2pt}
\centering
\resizebox{\columnwidth}{!}{
\begin{tabular}{|c|c|c|c|c|c|}
\hline

\textbf{Cyberattack} & \begin{tabular}{@{}c@{}} \textbf{Attacked} \\ \textbf{neurons} \end{tabular} &\textbf{Voltage} & \begin{tabular}{@{}c@{}} \textbf{Attack} \\ \textbf{start} \end{tabular} & \begin{tabular}{@{}c@{}} \textbf{Attack} \\ \textbf{duration} \end{tabular} \\ \hline

FLO & 100 & $\uparrow$ 40 mV & 50 ms & Instantaneous (1 ms) \\ \hline
JAM & 100 & -65 mV & 10 ms & 50 ms \\ \hline
SCA & 200 & $\uparrow$ 40 mV & 10 ms & Whole simulation \\ \hline
FOR & 200 & $\downarrow$ 40 mV & 10 ms & Whole simulation \\ \hline
SPO & 100 & Recorded voltages & 10 ms & 50 ms  \\ \hline
SYB & 100 & Opposite in range & 10 ms & Instantaneous (1 ms) \\ \hline
SIN & Up to 200 & $\uparrow$ 40 mV & 10 ms & Instantaneous (1 ms) \\ \hline
NON & Up to 200 & $\uparrow$ 40 mV, $\downarrow$ 40 mV, or 0 mV & 10 ms & Whole simulation  \\ \hline
\end{tabular}}
\end{table}

\begin{figure*}[h]
\begin{center}
\includegraphics[width=\textwidth]{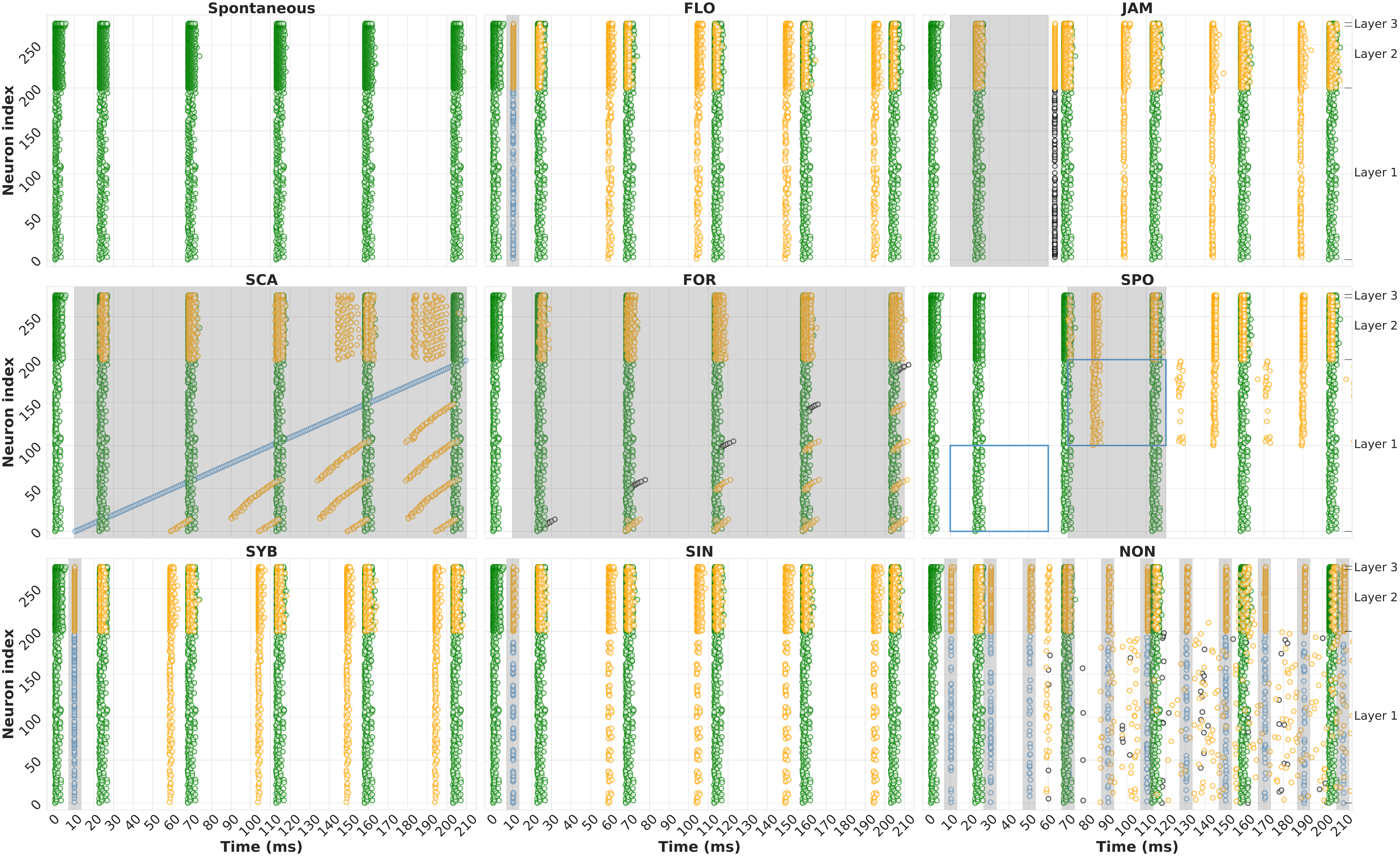}
\end{center}
\caption{Visual representation of the behavior of each neural cyberattack proposed. Green dots represent neuronal spikes from the spontaneous behavior, blue dots indicate stimulated neurons, black dots inhibited ones, and orange dots highlight the spikes produced as a consequence of the attack. A grey background indicates the duration of the attack.}
\label{fig:taxonomy_visual}
\end{figure*}

To better understand the behavior of these cyberattacks and the parameters indicated, \figurename~\ref{fig:taxonomy_visual} depicts a raster plot per cyberattack with the evolution of neuronal spikes during the first 215 ms of a complete simulation of 27 seconds. It allows the visual comparison between each cyberattack and the spontaneous behavior. 

As can be seen in \figurename~\ref{fig:taxonomy_visual}, the first raster plot, representing the spontaneous behavior, presents vertical columns of green dots corresponding to regular spiking from Izhikevich's model. This spontaneous behavior is also included in the plots presenting each cyberattack to compare their behavior easily. Blue dots indicate neurons attacked by neural stimulation, while black dots represent inhibitory actions. Furthermore, orange dots highlight the evolution of each cyberattack. Finally, a grey background indicates the duration of the cyberattack.

Compared to the spontaneous behavior, FLO generates new orange groups of spikes before the spontaneous columns, caused by the stimulation performed at 10 ms. Additionally, orange spikes can be appreciated within the green columns in layers two and three (neurons 200 to 276). These spikes are also a consequence of the attack, applying to subsequent cyberattacks. On the contrary, JAM performs neural inhibition until the instant 60 ms, and it is after that instant when the subset of attacked neurons performs spikes (indicated in black), inducing a delay compared to the spontaneous behavior that is repeated over time as a second column of orange spikes. 

Regarding SCA and FOR, both cyberattacks are active during almost all the simulation. However, their impact is quite different. In SCA, a diagonal succession of stimulated neurons can be observed, producing an incremental impact propagated along time. This impact can be appreciated by the apparition of additional diagonal groups of spikes under the diagonal and the anticipation of spikes in the second and third layers. In contrast, FOR only presents small perturbations compared to the spontaneous behavior induced by the implementation considerations already presented in \figurename~\ref{fig:flowchart_FOR}. Furthermore, SPO also performs its activity during a temporal window. In this case, there is a clear difference between the behavior of neurons with indexes 100 to 200 compared to the spontaneous behavior caused by the repetition of spikes previously recorded between instants 10 to 60 ms.

Moving to another stimulation cyberattack, SYB presents a similar spikes trend to FLO. This is explained by the voltage range defined by Izhikevich's model, between -65 mV to 30 mV, which introduces a higher probability of stimulating than inhibiting neurons. The instant of attack is also relevant since if a large population of neurons recently performed spikes, the voltage will be low and it will tend to induce stimulation actions. Although the output in terms of spikes is similar, their inner behavior is different.

SIN is another neural cyberattack that also presents similarities with FLO in terms of the visual distribution of spikes. However, it can be seen that there is a particular pattern in the attacked neurons, caused by the real target of the attack: neuron 201, the first neuron of the second layer. In this particular topology, it is determined by the connections between layers of the computational model used. Finally, NON induces a more chaotic behavior when the attack progresses, evaluating the attack condition every 20 ms. As can be seen, it performs both stimulation and inhibition tasks, randomly selected for each instant under attack and neuron of the first layer.

\begin{figure}[h]
\begin{center}
\includegraphics[width=0.9\columnwidth]{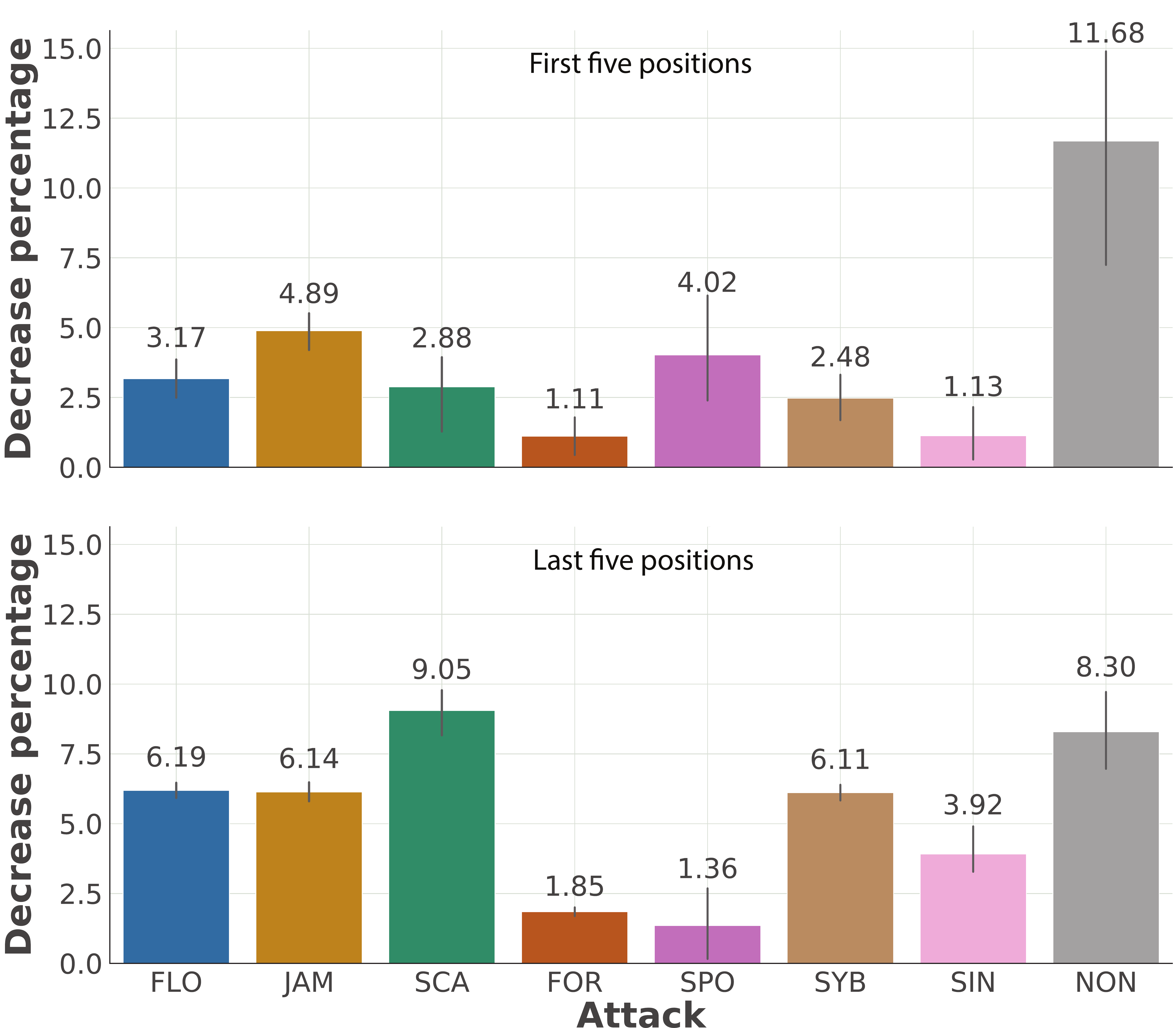}
\end{center}
\caption{Percentage of spikes reduced per neural cyberattack compared with spontaneous behavior, studied over the first and last five positions of the maze.}
\label{fig:results}
\end{figure}

Once introduced the behavior of each neural cyberattack graphically, \figurename~\ref{fig:results} depicts the impact caused by each cyberattack compared to spontaneous behavior, indicating the percentage of spikes reduction. This figure shows a differentiation between the first five positions and the last five positions of the optimal path of the maze to find the exit, determining which cyberattacks are more harmful in the short term and which are more suitable for long-term attacks. NON, due to its random behavior, achieves a reduction of almost 12\% over spontaneous activity in the first five positions, being the most damaging cyberattack in the short term, followed by JAM with almost a 5\% of reduction. In contrast, SCA is the most impacting attack for the long term, causing a spike reduction of around 9\%, followed by NON with a reduction of 8\%.

To conclude, it is essential to mention that the metric concerning the number of spikes has been selected due to its relevance on a wide variety of neuronal conditions. As an example, Amyotrophic Lateral Sclerosis (ALS) naturally generates hyperexcitability of neuronal activity. In this direction, a cyberattack based on neural stimulation could disrupt the natural equilibrium between neuronal excitation and inhibition, aggravating the status of the disease. Based on that, the number of spikes is considered an essential metric to evaluate the damage caused by a cyberattack.

%% file: tex/5conclusion.tex
Novel BCI generations bring countless benefits to society, improving their capabilities to offer better recording and stimulation resolutions. Moreover, the authors envision a future where the reduction in electrode size will derive in a broad coverage of the brain with single-neuron resolution. Although these improvements represent a paradigm change, vulnerabilities in these technologies open the door for cyberattacks to cause physical damage to users.

Based on the previous concerns, this work presents a taxonomy of eight neural cyberattacks aiming to disrupt spontaneous neural activity by maliciously inducing neuronal stimulation or inhibition, exploring the possibility of recreating the effects of particular neurodegenerative conditions. In this sense, two groups of cyberattacks are defined, either based on performing the attack at a particular instant or during a temporal window. These cyberattacks have been evaluated over a neuronal topology modeling a particular region of a mouse's visual cortex. Since there is a lack of realistic neuronal topologies nowadays, and following current literature, a Convolutional Neural Network has been trained to surpass this limitation due to their similarities with biological ones.

The impact of each cyberattack has been measured and compared over a common neural topology, being Neural Nonce and Neural Jamming the most damaging cyberattacks in the short term, causing a spike reduction of around 12\% and 5\% over spontaneous signaling, respectively. In contrast, Neural Scanning and Neural Nonce are more suitable for long-term damage, causing an approximate spike reduction of 9\% and 8\%, respectively.